\documentstyle[12pt,aaspp4,tighten]{article} 
\def\farcs{\hbox{$.\!\!^{\prime\prime}$}} 
 
\def\pagebreak{\vfill\eject}

\def\sun{\mbox{$_\odot$}}

\def\deg{{$^\circ$}}
\def\half{{\leavevmode\kern.1em\raise.5ex
\hbox{\the\scriptfont0 1}\kern-.1em /
\kern-.15em\lower.25ex\hbox{\the\scriptfont02}}} %exercise 11.6
\def\gtsim{\lower.5ex\hbox{$\buildrel > \over\sim$}}
\def\ltsim{\lower.5ex\hbox{$\buildrel < \over\sim$}}
\def\sun{\mbox{$_\odot$}}
\def\kms{~km~s$^{-1}$}

\def\hco{HCO$^+$}
\def\htco{H$^{13}$CO$^+$}
\def\h{H$_2$}

\def\water{H$_2$O}
\def\methanol{CH$_3$OH}

\def\jyb{~Jy~beam$^{-1}$}

\def\mjybkms{~mJy~beam$^{-1}$ km~s$^{-1}$}
\def\mjyb{~mJy~beam$^{-1}$}
\def\ra#1#2#3{#1$^{\rm h}$#2$^{\rm m}$#3$^{\rm s}$}
\def\dec#1#2#3{$#1^\circ#2'#3''$}
\def\arcs{\hbox{$^{\prime\prime}$}}
\def\mag{\hbox{$^{\rm m}$}}
\def\fmag{\hbox{$\,.\!\!^{\rm m}$}}
\slugcomment{15oct03, re-submitted to ApJ, referee comments incorporated}

\begin{document}

\title{
Discovery of a Massive Protostar near IRAS\,18507$+$0121}

\author{
D. S. Shepherd\altaffilmark{1}, 
D. E. A. N\"{u}rnberger\altaffilmark{2}, \& 
L. Bronfman\altaffilmark{3} 
}

\vspace{-3mm}
\altaffiltext{1}{National Radio Astronomy Laboratory, P.O. Box O, 1003
Lopezville Rd, Socorro, NM 87801}
%% \altaffiltext{2}{Chile}
%% \altaffiltext{3}{European Southern Observatory, Casilla 19001,
%% Santiago 19, Chile}
\altaffiltext{2}{European Southern Observatory, Casilla 19001, 
Santiago 19, Chile}
\altaffiltext{3}{Departamento de Astronom\'{\i}a, Universidad de Chile, 
Casilla 36-D, Santiago, Chile}

\rightskip=\leftskip 
\vspace{-3mm}

\begin{abstract}

We have observed the massive star forming region, IRAS\,18507$+$0121,
at millimeter wavelengths in 3~mm continuum emission and
{\htco}(J=1--0) and SiO(v=0, J=2--1) line emission, and at
near-infrared wavelengths between 1.2 and 2.1\,$\mu$m.  Two compact
molecular cores are detected: one north and one south separated by
$\sim 40''$.  The northern molecular core contains a newly discovered,
deeply embedded, B2 protostar surrounded by several hundred solar
masses of warm gas and dust, G34.4+0.23~MM.  Based on the presence of
warm dust emission and the lack of detection at near-infrared
wavelengths, we suggest that G34.4+0.23~MM may represent the
relatively rare discovery of a massive protostar (e.g. analogous to a
low-mass ``Class 0'' protostar).  The southern molecular core is
associated with a near-infrared cluster of young stars and an
ultracompact (UC) HII region, G34.4$+$0.23, with a central B0.5 star.
The fraction of near-infrared stars with excess infrared emission
indicative of circumstellar material is greater than 50\% which
suggests an upper limit on the age of the IRAS\,18507$+$0121 star
forming region of 3 Myrs.

\end{abstract}

\vspace{-.5cm}

\clearpage
%%%%%%%%%%%%%%%%%%%%%%%%%%%%%%%%%%%%%%%%%%%%%%%%%%%%%%%%%%%%%%%%%
%%%%%%%%%%%%%%%%%%%%%%%%%%%%%%%%%%%%%%%%%%%%%%%%%%%%%%%%%%%%%%%%%
\section{INTRODUCTION}

The massive star forming region associated with IRAS\,18507$+$0121
(hereafter IRAS\,18507) is
located 3.9~kpc from the Sun, and is roughly $11'$ from the HII region
complex G34.3+0.2 (Molinari et al. 1996, Carral \& Welch 1992).  Near
IRAS\,18507, Miralles et al. (1994) discovered an ultracompact
(UC) HII region (G34.4+0.23) embedded in a 1000~M\sun\ molecular cloud
traced by NH$_3$ emission.  The NH$_3$ emission is elongated in the
N-S direction with a total extent of about $7'$, however the $1.5'$
resolution of the observations was not adequate to discern the
structure of the core (Miralles et al. 1994).  The detection of
unresolved {\hco} and SiO emission (HPBW $55''$ and $43''$,
respectively) is reported by Richards et al. (1987) and Harju et
al. (1998).

IRAS\,18507 was detected in a CS(2-1) survey of IRAS point sources
with far-infrared colors suggestive of UC H II regions (Bronfman et
al. 1996).  The source was selected for further high resolution
studies because of its broad line wings, a signature of current star
formation.  By modeling {\hco}, {\htco}, CS and C$^{34}$S spectra
obtained at an angular resolution of $\sim\,16''$ Ramesh et al.\
(1997) demonstrate that the observed line profiles can be explained by
a collapsing hot core of about 800\,M$\sun$ which is hidden behind a
cold ($\sim$\,4\,K) and dense ($3 \times 10^{4}$\,cm$^{-3}$)
envelope of about 200\,M$\sun$.  The IRAS\,18507 region is also associated
with variable {\water} maser (Scalise et al. 1989; Palla et al. 1991;
Miralles et al. 1994) and {\methanol} maser emission (Schutte et
al. 1993; Szymczak et al. 2000).  Molinari et al. (1996, 1998)
observed IRAS\,18507 (labeled Mol74 in their papers) and
estimated a deconvolved size of the UC HII region of 0.7'' (0.013\,pc
at D=3.9\,kpc).

To date, the molecular gas and near-infrared emission have not been
observed with arcsec resolution.  Given the distance to the source of
nearly 4 kpc, the current low-resolution observations have not made it
possible to determine the evolutionary status of the region or the
relationship between the IRAS source, the UC HII region, and the
molecular gas.  In this work, we present observations of
IRAS\,18507 at near-infrared wavelengths, in millimeter
continuum emission tracing warm dust, and in the dense core tracer
{\htco}(J=1--0) and the shock tracer SiO(v=0, J=2--1) with $\sim 5''$
resolution.

%%%%%%%%%%%%%%%%%%%%%%%%%%%%%%%%%%%%%%%%%%%%%%%%%%%%%%%%%%%%%%%%%
%%%%%%%%%%%%%%%%%%%%%%%%%%%%%%%%%%%%%%%%%%%%%%%%%%%%%%%%%%%%%%%%%
\section{OBSERVATIONS}

%%%%%%%%%%%%%%%%%%%%%%%%%%%%%%%%%%%%%%%%%%%%%%%%%%%%%%%%%%%%%%%%%
\subsection{Observations in the 3~mm band}

Simultaneous observations in the 3 mm continuum band, and
{\htco}(J=1--0) and SiO(v=0, J=2--1) lines were made with the Owens
Valley Radio Observatory (OVRO) array of six 10.4~m telescopes on 1998
April 19 and 1998 May 17.  Projected baselines ranging from 12 to 120
meters provided sensitivity to structures up to about $20''$ with
$\sim 5''$ resolution.
% 1 each of L and E configuration tracks 
The total integration time on source was approximately 6.4 hours.
Cryogenically cooled SIS receivers operating at 4~K produced typical
single sideband system temperatures of about 400~K.  The gain
calibrator was the quasar $1749+096$ and the passband calibrators were
3C~454.3 and 3C~273.  Observations of Neptune provided the flux
density calibration scale with an estimated uncertainty of $\sim
15$\%.  Calibration was carried out using the Caltech MMA data
reduction package (Scoville et al. 1993).  Images were produced using
the MIRIAD software package (Sault, Teuben, \& Wright 1995) and
deconvolved with a CLEAN-based algorithm.

Continuum images with a 1~GHz bandwidth centered at frequency
89.983~GHz have a synthesized beam $4.9'' \times 4.2''$ (FWHM) at
P.A. $-31.2^\circ$ and RMS noise 3.3~mJy~beam$^{-1}$.  The spectral
band pass for {\htco} and SiO is centered on the systemic local
standard of rest velocity ($v_{LSR}$) $57.0$~\kms.  The {\htco} images
have a synthesized beam of $5.3'' \times 4.5''$ (FWHM) at
P.A. $-43.8$\deg, spectral resolution 1.728{\kms} and RMS noise
50.0\mjyb.  The SiO images have a synthesized beam of $5.3'' \times
4.5''$ (FWHM) at P.A. $-44.1$\deg\ and RMS noise 45{\mjyb} at a spectral
resolution 1.738~{\kms}.  SiO was not detected and the images are not
shown in this paper.

%%%%%%%%%%%%%%%%%%%%%%%%%%%%%%%%%%%%%%%%%%%%%%%%%%%%%%%%%%%%%%%%%
\subsection{Archival observations at 6~cm}

Archival data from the Very Large Array (VLA)\footnote{The National
Radio Astronomy Observatory is a facility of the National Science
Foundation operated under cooperative agreement by Associated
Universities, Inc.} were obtained at 4.88 GHz (6.15~cm) in continuum
emission with a bandpass of 100~MHz\footnote{Originally
published by Molinari et al. (1998) however, their image did not
include the position of the millimeter core.  Thus, we have re-calibrated
\& imaged their data to obtain a limit on the free-free continuum
emission toward the millimeter core.}.  Observations centered on 
IRAS\,18507 were made on 1994 October 4 with an on-source integration
time of about 4~minutes.  The absolute flux scale was derived from
observations of 3C~48 while the quasars 1801$+$010 and 1821$+$107 were
used as gain calibrators.  The data were calibrated and imaged using
the AIPS$++$ data reduction package.  The image had a synthesized beam
of $5.65'' \times 2.98''$ (FWHM) at P.A. $60.1^\circ$ and RMS noise
0.19~mJy~beam$^{-1}$.  Our resulting image is comparable to the
AIPS--generated image from Molinari et al. (1998).

%%%%%%%%%%%%%%%%%%%%%%%%%%%%%%%%%%%%%%%%%%%%%%%%%%%%%%%%%%%%%%%%%
\subsection{Near-infrared observations}

Broad band J, H and K$^{\prime}$ ($\lambda_{\rm
c}$\,$=$\,1.25\,$\mu$m, 1.65\,$\mu$m, and 2.10\,$\mu$m, respectively)
observations were performed in 1999 May at the Las Campanas 2.5\,m
Du\,Pont telescope using the facility NIR camera (Persson et al.\
1992) equipped with a NICMOS\,3 256\,$\times$\,256 HgCdTe array
detector.  The chosen plate scale of 0$\farcs$348 per pixel provides a
field of view of about
90$^{\prime\prime}$\,$\times$\,90$^{\prime\prime}$, thus covering an
area of 1.7\,pc\,$\times$\,1.7\,pc at the distance of IRAS\,18507 (3.9
kpc).

To remove randomly distributed cosmic rays we applied a dithering
method using $20''$ to $25''$ offsets.  In each filter
the exposure time was 10\,s per frame, resulting in an on-source 
integration time of up to 70\,s.  During acquisition of the data
set we had photo\-metric conditions throughout the night and the
seeing was 0$\farcs$6 to 0$\farcs$8 FWHM.  For photo\-metric
calibration we observed the faint NIR standard stars 9175 ($=$\,S071-D) 
and 9137 ($=$\,S372-S) from the sample of Persson et al.\ (1998). 

Following the procedure given by Persson et al.\ (1998), data
reduction (e.g. dark current subtraction, flat fielding, bad pixel
correction and sky subtraction) was performed using standard IRAF
software packages.  The Digitized Sky Surveys (1\,\&\,2) and the HST
Guide Star Catalog (both provided by the Space Telescope Science
Institute) were used to obtain astro\-metric calibration with an
accuracy better than $\pm$1$^{\prime\prime}$.  The detection limits in
%% the near-infrared images are J = 23 mag, H = 22 mag and K = 21.5 mag. 
%% COMMENT BY DIETER: I HAVE DISCARDED THE FAINTEST SOURCES IN ALL THREE 
%% FILTERS AS THEIR MAGNITUDE UNCERTAINTIES ARE REDICULOUSLY HIGH. 
%% I ONLY KEPT SOURCES WITH SIGMA VALUES SMALLER THAN 2.0 MAGNITUDES. 
the near-infrared images are J $=$ 19.6 mag, H $=$ 19.2 mag and
K$^{\prime}$ $=$ 18.6 mag.  Due to relatively high read-out noise the
quality of the photo\-metric calibration is restricted to $\sigma_{\rm
J}$ $=$ 1$\fmag$2\,$\pm$\,0$\fmag$5, $\sigma_{\rm H}$ $=$
1$\fmag$4\,$\pm$\,0$\fmag$6 and $\sigma_{\rm K}$ $=$
1$\fmag$6\,$\pm$\,0$\fmag$6.

%%%%%%%%%%%%%%%%%%%%%%%%%%%%%%%%%%%%%%%%%%%%%%%%%%%%%%%%%%%%%%%%%
%%%%%%%%%%%%%%%%%%%%%%%%%%%%%%%%%%%%%%%%%%%%%%%%%%%%%%%%%%%%%%%%%
\section{RESULTS}

A single, unresolved, newly discovered, millimeter continuum source,
G34.4$+$0.23~MM (hereafter G34.4~MM), is detected at $\alpha =$
\ra{18}{53}{18.01} ~ $\delta =$ \dec{+01}{25}{25.6}~(J2000) with a
total flux density of 56.8\,mJy (Fig. 1).  The 6~cm continuum image
showing free-free emission from ionized gas is also shown in Fig. 1
(Molinari et al. 1998).  Coincident with G34.4~MM is a marginal
detection of a 6~cm source with a flux density of 0.7\,mJy
(3.5\,$\sigma$).  The unresolved UC HII region, G34.4031$+$0.2276,
discovered by Molinari et al. (1998) is located about $40''$ south of
the millimeter core at $\alpha =$ \ra{18}{53}{18.68} ~ $\delta =$
\dec{+01}{24}{47.2}~(J2000) and has a flux density of 9.0\,mJy.

The total {\htco} flux density from 53.54{\kms} to 62.16\kms\ is
33.64~Jy (Fig. 2).  The strongest {\htco} peak is located
$\sim 3''$ northwest of the G34.4 UC HII region.  A second {\htco}
peak is coincident with G34.4~MM.  The two peaks are connected by a
band of more diffuse molecular gas traced by {\htco}.  The total
extent of the dense {\htco} gas is consistent with previous NH$_3$
observations made at $1.5'$ resolution (Miralles et al. 1994).

SiO(J=2--1) emission is not detected with a $3 \sigma$ upper limit of
135{\mjyb}.  This is in contrast to the single dish SiO spectrum
($43''$ resolution) obtained by Harju et al. (1998) in which they
detect a $5 \sigma$ peak intensity of 0.26\,K ($\sim
840$\,mJy). However, this discrepancy could be explained if the SiO
emission is spatially extended and thus missed by our interferometric
observations.  Indeed, observations of IRAS\,18507 at $16''$ resolution with
the Nobeyama 45\,m telescope shows extended SiO emission with a FWHM
of about 1 arcmin (Bronfman, personal communication).  At the position
of IRAS\,18507 the spectrum is non-Gaussian with a peak
temperature of $\sim 0.7$\,K (1.1{\jyb}) and a total measured velocity
extent at zero intensity of $\sim 15${\kms}.  Assuming the peak
emission scales with the area of the synthesized beam, we would expect
to detect a peak flux density of roughly 110{\mjyb} ($2.4 \sigma$)
with our $5''$ resolution.  However, our interferometric observations
are only sensitive to structures up to $\sim 20''$ and the actual
source size is nearly 1 arcmin FWHM. Thus, our lack of detection in
SiO(J=2--1) is likely due to a combination of the interferometer
missing zero-spacing flux and low sensitivity to any remaining compact
emission.

Near-infrared images of the region are shown in Fig. 3 while a
comparison between the 2.10\,$\mu$m emission and the {\htco} emission
is shown in Fig. 4.  A near-infrared cluster of young stars is located
on the western edge of the southern {\htco} core.  The brightest
member of the cluster (labeled as source $\#$54 in Figs.  3 \& 4) has
near-infrared magnitudes: J $=$ 17.0$\pm$0.6 mag, H $=$ 15.3$\pm$0.5
mag, and K$^{\prime}$ $=$ 14.2$\pm$0.4 mag.  The G34.4 UC HII region
appears to be a member of this young stellar cluster although it is
not detected at 2.10\,$\mu$m.  In contrast, the northern {\htco} core
does not have an associated stellar cluster nor is the exciting star
of G34.4~MM detected in the near-infrared.

In Table 1 we summarize the results of the JHK$^{\prime}$ imaging.
For all detected sources we list in columns 2 and 3 the positional
offsets (in arcsec) relative to the position of the brightest member
of the cluster ($=$\,$\#$54) as well as in columns 4 and 5 the
absolute coordinates.  Finally, columns 6-8 contain the J, H and
K$^{\prime}$ band magnitudes.  As already mentioned above, typical
photometric errors are in the range
$\sigma_{\rm J}$\,$=$\,1$\fmag$2\,$\pm$\,0$\fmag$5, 
$\sigma_{\rm H}$\,$=$\,1$\fmag$4\,$\pm$\,0$\fmag$6 and 
$\sigma_{\rm K}$\,$=$\,1$\fmag$6\,$\pm$\,0$\fmag$6.  
In total we have detected 146 sources, however, 80 of them are only
seen in the K$^{\prime}$ data which suggests that the extinction is so
large that many cluster members are too heavily extincted to be
detectable at wavelengths shorter than $2 \mu$m.  43 sources are
detected in all three filters, which enabled us to calculate
corresponding (J$-$H), (H$-$K$^{\prime}$) and (J$-$K$^{\prime}$)
colors.  Assuming a distance of 3.9\,kpc but not taking into account
any possible foreground extinction, we also derive lower limits for
the absolute magnitudes M$_{\rm J}$, M$_{\rm H}$ and M$_{\rm
K^{\prime}}$.

The NIR colors and absolute magnitudes allow us to place the sources
in near-infrared two-color and color-magnitude diagrams (see Figs. 5
and 6).  In both diagrams the location of source $\#$54 is indicated
by an asterisk\footnote{Note: source $\#$54 is shown in the figures
only to give a reference point to the brightest NIR member of the
cluster and to allow the reader to relate positional offsets from
$\#$54 given in Table 1 to other sources in the field.}  and, using
the extinction transformations given by Rieke \& Lebofsky (1985), a
reddening vector for A$_{\rm V}$\,$=$\,5\,mag is given.  In the
(H$-$K$^{\prime}$)--(J$-$H) diagram the loci of the observed sources
are compared to those of dwarfs and giants (thick and thin line,
respectively; Koornneef 1983).  Similarly, in the M$_{\rm
J}$--(J$-$K$^{\prime}$) diagram we mark the loci of stars on the main
sequence (thick line) and on the pre-main sequence at ages 0.3\,Myr
and 3\,Myr (two dashed lines; Palla \& Stahler 1993).

Following Strom et al. (1993), low mass stars possessing near-infrared
excess emission usually populate three distinct regions of the
(H$-$K$^{\prime}$)--(J$-$H) diagram: sources associated with low
excess emission (so-called weak-line T\,Tauri stars) are found in zone
I, while zones II and III feature sources with high excess emission
caused by circumstellar disks (classical T\,Tauri stars) and
surrounding envelopes (protostars), respectively.  The same or at
least a similar scenario probably holds for intermediate and high mass
stars (N\"urnberger 2003).  For about 50\,$\%$ of our sources with
JHK$^{\prime}$ detections we find the near-infrared colors clearly
offset from those of main sequence stars, e.g.\ source $\#$54 is
extincted by almost A$_{\rm V}$\,$=$\,20$\mag$.  Their reddening might
be convincingly explained by intrinsic extinction due to the presence
of relatively large amounts of circumstellar gas and dust.  The other
50\,$\%$ of our JHK$^{\prime}$ detected sources have colours which are
reddened by at most 2--3 magnitudes (likely due to foreground
extinction) and, apart from that, appear to be consistent with those
of main sequence stars.

%%%%%%%%%%%%%%%%%%%%%%%%%%%%%%%%%%%%%%%%%%%%%%%%%%%%%%%%%%%%%%%%%
%%%%%%%%%%%%%%%%%%%%%%%%%%%%%%%%%%%%%%%%%%%%%%%%%%%%%%%%%%%%%%%%%
\section{Discussion}

%%%%%%%%%%%%%%%%%%%%%%%%%%%%%%%%%%%%%%%%%%%%%%%%%%%%%%%%%%%%%%%%%
\subsection{Ionized gas emission}

The UC HII region G34.4$+$0.23 is detected at 6~cm with a flux density
of 9$\pm 0.2$\,mJy while G34.4~MM has an unresolved 0.7$\pm 0.2$\,mJy
6~cm continuum peak associated with it.  Thus, there are one or more
early-type stars producing ionized gas toward both sources.  Following
the method outlined by Wood \& Churchwell (1989), the physical
properties of the ionized gas are calculated and presented in Table 2.
For each source, the values listed are: $S_{\nu}$, the measured flux
density at 4.8851\,GHz; $\Delta s$, line-of-sight depth at the peak
position (an upper limit for unresolved sources); T$_b$, the
synthesized beam brightness temperature; $\tau_\nu$, the optical depth
assuming the beam is uniformly filled with T$_e = 10^4$ K ionized gas;
EM, the emission measure; $n_e$, the RMS electron density; U, the
excitation parameter of the ionized gas; $N_L$, the number of Lyman
continuum photons required to produce the observed emission assuming
an ionization-bounded, spherically symmetric, homogeneous HII region;
and finally, the spectral type of the central star assuming a single
ZAMS star is producing the observed Lyman continuum flux (Panagia
1973).  These estimates should be considered a lower limit if there is
significant dust absorption within the HII region or if the emission
is being quenched by high accretion rates expected for OB protostars
(e.g. Churchwell 1999 and references therein).

The values listed in Table 2 do not take into account possible dust
absorption within the ionized gas, which would tend to underestimate
$N_L$, and hence the spectral type of the star.  Estimates for
G34.4\,MM could also be uncertain by up to 50\% due to the low
signal-to-noise detection at 6~cm (3.5 $\sigma$).  Despite these
uncertainties, the derivations are probably accurate to within a
spectral type.  Comparison of the values in Table 2
with those in Wood \& Churchwell (1989, their Table 17), shows that
the physical parameters of the ionized gas in the G34.4 sources are
consistent with ZAMS stars with spectral types later than B0.  Values
for the UC HII region G34.4$+$0.23 are consistent with those derived
by Molinari et al. (1998) to within the errors.

%%%%%%%%%%%%%%%%%%%%%%%%%%%%%%%%%%%%%%%%%%%%%%%%%%%%%%%%%%%%%%%%%
\subsection{Thermal dust emission at millimeter wavelengths}

The UC HII region G34.4$+$0.23 has no detectable millimeter continuum
emission coincident with the ionized gas peak.  Further, members of
the near-infrared cluster also show no millimeter continuum emission
coincident with the stellar positions.  The $3 \sigma$ upper limit on
warm dust emission is $\sim 10${\mjyb}.  The mass of gas and dust is
estimated from the millimeter continuum emission using ${\rm M}_{gas +
dust} = \frac{{\rm F}_{\nu}~ {\rm D}^2} {{\rm B}_{\nu}({\rm T}_d)~
\kappa_{\nu}}$ where D is the distance to the source, ${\rm F}_{\nu}$
is the continuum flux density due to thermal dust emission at
frequency $\nu$, ${\rm B}_{\nu}$ is the Planck function at temperature
T$_d$ (Hildebrand 1983).  The dust opacity per gram of gas is
estimated from $\kappa_{\nu} = 0.006(\frac{\nu}{245 {\rm
GHz}})^{\beta}$~cm$^2$~g$^{-1}$ where $\beta$ is the opacity index
(see Kramer et al. 1998; and the discussion in Shepherd \& Watson
2002).  We assume the emission is optically thin, and the temperature
of the dust can be characterized by a single value.  We take T$_d$ to
be 50~K based on measurements of typical conditions in warm molecular
cores with embedded protostars (Hogerheijde et al. 1998) and $\beta =
1.5$ (Pollack et al. 1994).  We also assume a distance of 3.9~kpc
(Molinari et al. 1996) and a gas-to-dust ratio of 100 (Hildebrand
1983).  Thus, the upper limit to the mass of gas and dust that can
exist around the UC HII region and still be below our detection
threshold is 40\,M{\sun}.  

G34.4\,MM has a strong millimeter continuum peak.  Assuming
the ionized gas is optically thin between 6~cm and 3~mm ($S_{\nu}
\propto \nu^{-0.1}$), we expect a contribution of 0.52~mJy to the flux
density at 3~mm.  Thus, the total flux density due to thermal dust
emission at 3~mm is 56.3~mJy.  We find the mass of gas and dust
associated with thermal dust emission at 3~mm is approximately
240~M{\sun}.  Changing the assumptions of T$_d$ and $\beta$, we find
mass estimates vary from 150~M{\sun} with T$_d$ = 50~K \& $\beta = 1$
to as high as 650~M{\sun} with T$_d$ = 30~K \& $\beta = 2$.  Despite
the uncertainties associated with this estimate, our results show that
there are several hundred solar masses of warm gas and dust in this
region.  Assuming the G34.4~MM core is heated internally, then this
large molecular mass is consistent with the presence of a massive,
embedded OB star or a cluster of massive stars (e.g. Saraceno et
al. 1996).
% mass of circumstellar material is proportional to the bolometric
% luminosity of the embedded protostar.

%%%%%%%%%%%%%%%%%%%%%%%%%%%%%%%%%%%%%%%%%%%%%%%%%%%%%%%%%%%%%%%%%
\subsection{{\htco} emission}

To estimate the average column density and mass associated with the
{\htco} emission (Table~3) we assume the {\htco} is optically thin and
the rotational temperature follows the average kinetic temperature
derived from NH$_3$ observations, e.g. $T_{rot} = 22$~K (Molinari et
al. 1996).  Temperatures are likely to be higher in cores with
embedded sources and lower in the diffuse gas however, an average
temperature of 22\,K should be a reasonable estimate.  Typical
uncertainties are a factor of 2--3.

Abundances of [\hco]/[\h] in massive star forming regions are
typically $\sim 10^{-9}$ (Blake et al. 1987).  Assuming an isotopic
ratio [\hco]/[\htco] $\sim 51$ for the galacto-centric distance to
IRAS\,18507 of 5.8~kpc (Wilson \& Rood 1994), we derive a total mass
of the {\htco} cloud to be $5 \times 10^{4}$~M\sun\ and compact core
masses of $4-5 \times 10^{3}$~M\sun.  This estimate can easily be off
by an order of magnitude if \hco, and hence \htco\ is enhanced due to
shocks.  Comparing our mass estimates with that derived from NH$_3$ of
1000\,M\sun, and assuming that the {\htco} and NH$_3$ trace the same
volume of gas, our estimates are an order of magnitude larger
suggesting that some enhancement of the {\hco} abundance has likely
occurred in this region.

With {\htco} column densities in hand we can attempt to constrain the
intrinsic extinction of the central sources of both molecular cores.
Taking into account abundances and isotopic ratios as given above, the
{\htco} column densities of 2\,$\times$\,10$^{14}$\,cm$^{-2}$ convert
into N({\h})\,$\sim 10^{25}$\,cm$^{-2}$.  Following
Bohlin et al. (1978), {\h} column densities and A$_{\rm V}$ are
related via the formula N({\h})\,/\,A$_{\rm V}$ $=$
0.94\,$\times$\,10$^{21}$\,cm$^{-2}$\,mag$^{-1}$, for A$_{\rm
V}$\,$<$\,1\,mag.  This suggests intrinsic extinctions of the order
10$^{4}$\,mag toward the central sources of the two
G34.4 cores.  We emphasize that these A$_{\rm V}$ values represent
only rough estimates because the given N({\h})\,/\,A$_{\rm V}$
relation might flatten significantly for A$_{\rm V}$\,$\gg$\,1\,mag
(see Dickman 1978 and Frerking et al. 1982).  Nevertheless, such large
intrinsic extinctions easily explain why no near-infrared sources are
detected toward the G34.4\,MM core.  Similarly, the near-infrared
sources seen in the neighborhood of the southern core are probably
located at its periphery and not at its center.

%%%%%%%%%%%%%%%%%%%%%%%%%%%%%%%%%%%%%%%%%%%%%%%%%%%%%%%%%%%%%%%%%
\subsection{Circumstellar material around members of the near-infrared 
cluster}

As discussed in Section 3, Figs. 5 \& 6 show that about 50\% of
cluster members with JHK$'$ detections have near-infrared colors
clearly offset from those of main sequence stars suggesting the
presence of circumstellar gas and dust (albeit below our millimeter
continuum detection limit).  Observational evidence for the presence
of disks in clusters of varying ages suggests that in low- and
intermediate-mass star forming regions half of all stars loose their
disks within 3 Myrs and 90\% of stars loose their disks within 5 Myrs
(e.g. Robberto et al. 1999; Meyer \& Beckwith, 2000; Haisch, Lada, \&
Lada 2001).  Low mass dust disks (as low as 0.1 lunar masses) may even
persist as long as a billion years (Spangler et al. 2001).

Only 30\% of the members of the NIR cluster are detected in all
three bands, and half of those appear to have circumstellar material.
Using the relation between the percent of sources with NIR excess in a
cluster versus the age of the cluster (Haisch, Lada, \& Lada 2001),
our JHK$'$ data suggest that the NIR cluster is less than 3 Myrs
old.  This is consistent with the large number of sources seen in Fig.
6 which have (J--K$'$) colors well in excess of the 3 Myr pre-main
sequence locus.  How does this estimate compare with estimates of disk
dispersal times?

The most massive star in the NIR cluster (source $\#$54) has an
infrared excess suggesting that it still has circumstellar material.
Assuming the material resides in a disk, how long would it take for
this star to photoevaporate its disk?  Using the `` weak wind'' model
of Hollenbach et al. (1994), the lifetime of a circumstellar disk is
given by:
\begin{equation}
\tau_{disk} = 7 \times 10^4 ~\Phi_{49}^{-1/2} ~M_1^{-1/2}~M_d ~~{\rm [yrs]}
\end{equation}
where\\
\begin{tabular}{lll}
~~~ & $\Phi_{49}$ & = ionizing Lyman continuum flux in units of
    $10^{49}$~s$^{-1}$  \\
    & $M_1$ & = the mass of the central star in units of 10~M\sun \\
    & $M_d$ & = disk mass in units of M\sun \\
\end{tabular}
\\ For source $\#$54 we assume M$_\star = 5$M{\sun}, $\Phi_{49} \sim 8
\times 10^{-7}$ (Thompson 1984) and $M_1 \sim 0.5$.  Shu et al. (1990)
showed that an accretion disk becomes gravitationally unstable when it
reaches a mass of $M_d \sim 0.3 M_\star$ where $M_\star$ is the mass
of the central protostar.  During the initial collapse of the cloud
core, the disk mass may be maintained close to the value of $0.3
M_\star$.  When infall ceases and the disk mass falls below the
critical value, disk accretion onto the star may rapidly decline and
photoevaporation may be the dominant mechanism which disperses the
remaining gas and dust (Hollenbach et al. 1994).  Based on this
scenario, we assume an initial disk at the edge of stability, that is
$M_d \sim 0.3 M_\star = 1.5 M$\sun.  Errors in the estimate for the
photoevaporative timescale would scale directly as $M_d$.  We find
that $\tau_{disk} \sim 10^8$ years.  For the less luminous stars in
the cluster, the photoevaporative timescale would be significantly
longer.  Thus, circumstellar disks should persist in the IRAS\,18507
star forming region for at least $10^8$ years.

%%%%%%%%%%%%%%%%%%%%%%%%%%%%%%%%%%%%%%%%%%%%%%%%%%%%%%%%%%%%%%%%%
%%%%%%%%%%%%%%%%%%%%%%%%%%%%%%%%%%%%%%%%%%%%%%%%%%%%%%%%%%%%%%%%%
\section{Summary}

Two massive molecular cores are detected toward IRAS\,18507.  The northern
molecular core is associated with a newly discovered, millimeter
continuum peak, G34.4~MM, produced by a mixture of thermal dust
emission (56.3~mJy), and ionized gas emission (0.52~mJy).  Several
hundred solar masses of warm gas and dust surround the central B2
star.  The high mass of warm circumstellar material is consistent with
the lack of detection at near-infrared wavelengths and suggests that
the source is, perhaps, younger than those near the southern molecular
core and may represent the relatively rare case of a massive protostar
(e.g. analogous to low-mass ``Class 0'' protostars).  If the central
protostar is undergoing accretion typical of early B protostars (e.g.
$\dot{M}_{acc} \sim 10^{-5}$ to $10^{-3}$), then there may be
significant dust absorption within the HII region or the emission
could be quenched by high accretion rates, both conditions would tend
to underestimate the spectral type of the embedded protostar
(e.g. Churchwell 1999 and references therein).

The southern core is associated with a near-infrared cluster of young
stars and a UC HII region, G34.4$+$0.23, with a central B0.5 star.  No
millimeter continuum emission is detected toward the peak of the UC
HII region suggesting that the source has had time to disperse a
significant fraction of the warm gas and dust which typically
enshrouds protostars during their early evolutionary phase.  The
molecular core has not been destroyed by the forming stars; instead
the stars appear to have formed around the periphery of the core,
leaving the core itself intact.  Further observations searching for
outflowing gas at millimeter wavelengths toward both molecular cores
would help constrain the relative age of the sources in IRAS\,18507.

Only 30\% of stars in near-infrared stellar cluster are seen in all
NIR bands.  Of those, 50\% have excess emission suggesting the
presence of at least some circumstellar material (well below our
millimeter continuum detection limit).  Based on the fraction of stars
with NIR excess, we estimate that the cluster is less than 3
Myrs old.

\noindent {\bf Acknowledgments}\\ 
Research at the Owens Valley Radio Observatory is supported by the
National Science Foundation through NSF grant number AST 96-13717.
Star formation research at Owens Valley is also supported by NASA's
Origins of Solar Systems program, Grant NAGW-4030, and by the Norris
Planetary Origins Project.  LB acknowledges support by FONDECYT Grant
1010431 and by Centro de Astrofisica FONDAP 15010003.

\clearpage
%table 1:

\small

\begin{table}[h!t]
  \begin{center}
    \begin{minipage}{16.0cm}
    \caption{\label{T:RESULTS-NIR} 
       Positions and JHK$^{\prime}$ photometry of all detected sources. 
       Sources $\#$54, 41, 43, 47, 49, 50, 53 
       and 55 are likely cluster members. }
    \begin{center}
%   \begin{tabular}{|r|*{2}{r|}*{5}{c|}}
    \begin{tabular}{|rrrccccc|}
      \hline
 SNB  & $\Delta$x & $\Delta$y &    RA   &   DEC   &   J   &   H   & K$^{\prime}$ \\
 $\#$ & [$\arcs$] & [$\arcs$] & (J2000) & (J2000) & [mag] & [mag] & [mag] \\
      \hline\hline
   1 & $-$14.4 & $-$33.7 & 18:53:17.48 & $+$01:24:18.1 & 16.5 & 15.8 & 15.6 \\
   2 &    40.2 & $-$32.9 & 18:53:21.12 & $+$01:24:18.9 &      &      & 18.1 \\
   3 &     7.5 & $-$32.7 & 18:53:18.94 & $+$01:24:19.1 &      &      & 18.1 \\
   4 &     3.9 & $-$32.5 & 18:53:18.70 & $+$01:24:19.3 &      &      & 15.9 \\
   5 &     5.3 & $-$32.3 & 18:53:18.79 & $+$01:24:19.5 &      &      & 15.8 \\
   6 & $-$12.4 & $-$31.9 & 18:53:17.61 & $+$01:24:19.9 & 18.2 & 17.3 & 16.6 \\
   7 &    24.2 & $-$31.9 & 18:53:20.06 & $+$01:24:19.9 &      &      & 18.2 \\
   8 &  $-$2.5 & $-$31.7 & 18:53:18.27 & $+$01:24:20.0 &      &      & 18.3 \\
   9 &    38.0 & $-$31.7 & 18:53:20.98 & $+$01:24:20.1 & 18.8 & 18.0 &      \\
  10 &    20.9 & $-$31.6 & 18:53:19.84 & $+$01:24:20.2 &      &      & 18.0 \\
  11 &    40.0 & $-$31.1 & 18:53:21.11 & $+$01:24:20.7 &      &      & 18.4 \\
  12 &    35.9 & $-$29.6 & 18:53:20.84 & $+$01:24:22.2 & 18.4 & 17.5 &      \\
  13 & $-$11.3 & $-$29.2 & 18:53:17.69 & $+$01:24:22.5 &      &      & 18.4 \\
  14 &    28.8 & $-$27.9 & 18:53:20.36 & $+$01:24:23.9 &      &      & 18.2 \\
  15 & $-$36.1 & $-$27.8 & 18:53:16.04 & $+$01:24:24.0 &      &      & 17.7 \\
  16 &     4.6 & $-$26.6 & 18:53:18.75 & $+$01:24:25.2 &      &      & 18.3 \\
  17 &    32.4 & $-$26.1 & 18:53:20.61 & $+$01:24:25.7 & 18.5 & 17.9 &      \\
  18 & $-$34.4 & $-$26.1 & 18:53:16.15 & $+$01:24:25.7 & 16.8 & 16.5 & 16.2 \\
  19 &    22.1 & $-$25.7 & 18:53:19.91 & $+$01:24:26.1 & 19.4 & 18.6 & 17.6 \\
  20 &    39.5 & $-$24.7 & 18:53:21.07 & $+$01:24:27.1 &      & 18.3 & 15.8 \\
  21 &  $-$6.7 & $-$24.5 & 18:53:18.00 & $+$01:24:27.3 &      & 18.2 & 17.0 \\
  22 &    30.2 & $-$24.2 & 18:53:20.45 & $+$01:24:27.6 & 18.9 & 18.3 & 17.9 \\
  23 &    30.0 & $-$23.7 & 18:53:20.44 & $+$01:24:28.1 & 19.5 &      &      \\
  24 & $-$37.1 & $-$22.9 & 18:53:15.96 & $+$01:24:28.9 &      &      & 16.4 \\
  25 &    15.3 & $-$21.5 & 18:53:19.46 & $+$01:24:30.3 & 19.5 & 18.2 & 17.5 \\
  26 &    11.6 & $-$20.3 & 18:53:19.21 & $+$01:24:31.5 &      & 18.9 & 18.0 \\
  27 & $-$40.6 & $-$19.5 & 18:53:15.73 & $+$01:24:32.3 &      &      & 17.8 \\
  28 & $-$10.9 & $-$18.3 & 18:53:17.71 & $+$01:24:33.5 &      &      & 18.5 \\
  29 & $-$18.5 & $-$16.8 & 18:53:17.21 & $+$01:24:34.9 &      & 18.6 & 17.7 \\
  30 &     6.8 & $-$15.6 & 18:53:18.90 & $+$01:24:36.2 &      &      & 18.4 \\
  31 &    35.0 & $-$14.8 & 18:53:20.77 & $+$01:24:37.0 & 14.5 & 14.3 & 14.1 \\
  32 & $-$26.0 & $-$14.7 & 18:53:16.71 & $+$01:24:37.1 & 17.9 & 16.4 & 15.7 \\
  33 &    37.8 & $-$13.1 & 18:53:20.96 & $+$01:24:38.7 & 17.1 & 16.2 & 15.7 \\
      \hline
    \end{tabular}
    \end{center}
    \end{minipage}
  \end{center}
\end{table}

\addtocounter{table}{-1}
\clearpage
\begin{table}[h!t]
  \begin{center}
    \begin{minipage}{16.0cm}
    \caption{\label{T:RESULTS-NIR-2} Continued. }
    \begin{center}
%   \begin{tabular}{|r|*{2}{r|}*{5}{c|}}
    \begin{tabular}{|rrrccccc|}
      \hline
 SNB  & $\Delta$x & $\Delta$y &    RA   &   DEC   &   J   &   H   & K$^{\prime}$ \\
 $\#$ & [$\arcs$] & [$\arcs$] & (J2000) & (J2000) & [mag] & [mag] & [mag] \\
      \hline\hline
  34 &  $-$7.4 & $-$12.8 & 18:53:17.95 & $+$01:24:39.0 &      &      & 18.3 \\
  35 & $-$37.2 & $-$12.1 & 18:53:15.96 & $+$01:24:39.7 & 18.0 & 16.7 & 15.9 \\
  36 &    38.5 & $-$11.3 & 18:53:21.00 & $+$01:24:40.5 &      &      & 16.8 \\
  37 &    32.5 & $-$10.1 & 18:53:20.61 & $+$01:24:41.7 &      &      & 17.8 \\
  38 &  $-$9.5 &  $-$8.4 & 18:53:17.80 & $+$01:24:43.4 &      &      & 18.2 \\
  39 &    19.0 &  $-$8.2 & 18:53:19.71 & $+$01:24:43.6 & 18.9 & 18.2 & 17.9 \\
  40 &  $-$1.0 &  $-$7.2 & 18:53:18.37 & $+$01:24:44.6 &      &      & 17.0 \\
  41 &  $-$0.6 &  $-$6.1 & 18:53:18.41 & $+$01:24:45.7 &      & 18.3 & 16.5 \\
  42 & $-$41.1 &  $-$5.7 & 18:53:15.70 & $+$01:24:46.0 & 17.9 & 16.8 & 16.2 \\
  43 &     0.7 &  $-$4.9 & 18:53:18.48 & $+$01:24:46.8 &      &      & 15.8 \\
  44 &    34.8 &  $-$4.6 & 18:53:20.76 & $+$01:24:47.2 &      & 18.7 & 17.9 \\
  45 & $-$26.6 &  $-$4.2 & 18:53:16.66 & $+$01:24:47.6 &      &      & 16.7 \\
  46 &    32.5 &  $-$3.9 & 18:53:20.61 & $+$01:24:47.9 &      &      & 16.9 \\
  47 &     2.0 &  $-$3.5 & 18:53:18.58 & $+$01:24:48.2 &      &      & 15.4 \\
  48 &    14.2 &  $-$3.1 & 18:53:19.40 & $+$01:24:48.7 & 19.0 & 18.3 & 18.3 \\
  49 &  $-$0.2 &  $-$2.1 & 18:53:18.43 & $+$01:24:49.7 &      &      & 15.2 \\
  50 &     6.6 &  $-$2.1 & 18:53:18.88 & $+$01:24:49.7 & 19.2 & 18.2 & 17.5 \\
  51 & $-$10.9 &  $-$1.9 & 18:53:17.71 & $+$01:24:49.9 &      &      & 18.1 \\
  52 &    34.9 &  $-$0.7 & 18:53:20.77 & $+$01:24:51.0 &      &      & 17.5 \\
  53 &  $-$3.1 &  $-$0.3 & 18:53:18.24 & $+$01:24:51.5 &      &      & 17.1 \\
  54 &     0.0 &     0.0 & 18:53:18.44 & $+$01:24:51.8 & 17.0 & 15.3 & 14.2 \\
  55 &  $-$4.3 &     1.9 & 18:53:18.15 & $+$01:24:53.7 &      &      & 18.0 \\
  56 &    23.2 &     2.2 & 18:53:19.99 & $+$01:24:54.0 & 17.3 & 16.7 & 16.3 \\
  57 & $-$11.2 &     2.7 & 18:53:17.69 & $+$01:24:54.5 & 18.6 & 17.5 & 16.7 \\
  58 & $-$35.1 &     2.8 & 18:53:16.11 & $+$01:24:54.6 &      &      & 17.6 \\
  59 &  $-$8.3 &     3.5 & 18:53:17.89 & $+$01:24:55.3 &      &      & 17.8 \\
  60 &     9.5 &     4.8 & 18:53:19.08 & $+$01:24:56.5 &      &      & 17.6 \\
  61 & $-$13.5 &     5.1 & 18:53:17.54 & $+$01:24:56.9 &      &      & 18.4 \\
  62 &    24.5 &     5.1 & 18:53:20.08 & $+$01:24:56.9 &      & 18.7 & 18.0 \\
  63 &    18.0 &     5.2 & 18:53:19.64 & $+$01:24:56.9 &      &      & 18.5 \\
  64 & $-$35.0 &     6.7 & 18:53:16.11 & $+$01:24:58.5 &      &      & 18.5 \\
  65 & $-$20.9 &     7.2 & 18:53:17.05 & $+$01:24:59.0 &      &      & 18.0 \\
  66 &     7.7 &     7.7 & 18:53:18.96 & $+$01:24:59.5 &      &      & 18.6 \\
      \hline
    \end{tabular}
    \end{center}
    \end{minipage}
  \end{center}
\end{table}

\addtocounter{table}{-1}
\clearpage
\begin{table}[h!t]
  \begin{center}
    \begin{minipage}{16.0cm}
    \caption{\label{T:RESULTS-NIR-3} Continued. }
    \begin{center}
%   \begin{tabular}{|r|*{2}{r|}*{5}{c|}}
    \begin{tabular}{|rrrccccc|}
      \hline
 SNB  & $\Delta$x & $\Delta$y &    RA   &   DEC   &   J   &   H   & K$^{\prime}$ \\
 $\#$ & [$\arcs$] & [$\arcs$] & (J2000) & (J2000) & [mag] & [mag] & [mag] \\
      \hline\hline
  67 &    21.5 &     8.1 & 18:53:19.88 & $+$01:24:59.9 & 17.7 & 17.2 & 16.9 \\
  68 &    35.4 &     8.8 & 18:53:20.80 & $+$01:25:00.6 & 18.2 & 17.4 & 17.1 \\
  69 &  $-$9.9 &     9.1 & 18:53:17.78 & $+$01:25:00.9 &      &      & 17.1 \\
  70 & $-$32.4 &     9.3 & 18:53:16.28 & $+$01:25:01.1 &      &      & 18.0 \\
  71 & $-$12.8 &     9.7 & 18:53:17.59 & $+$01:25:01.5 & 17.5 & 17.1 & 16.8 \\
  72 &     2.9 &    10.1 & 18:53:18.64 & $+$01:25:01.9 &      &      & 18.4 \\
  73 & $-$31.4 &    10.5 & 18:53:16.35 & $+$01:25:02.3 &      &      & 17.9 \\
  74 &  $-$9.2 &    10.6 & 18:53:17.83 & $+$01:25:02.4 &      &      & 17.1 \\
  75 &  $-$4.8 &    11.1 & 18:53:18.12 & $+$01:25:02.9 & 18.1 & 17.5 &      \\
  76 &    31.7 &    11.4 & 18:53:20.56 & $+$01:25:03.2 & 19.0 & 18.1 & 17.5 \\
  77 &  $-$6.4 &    11.6 & 18:53:18.01 & $+$01:25:03.4 &      &      & 18.2 \\
  78 & $-$14.9 &    11.7 & 18:53:17.45 & $+$01:25:03.5 &      &      & 18.5 \\
  79 &  $-$6.5 &    12.8 & 18:53:18.01 & $+$01:25:04.6 & 18.4 &      &      \\
  80 &    36.4 &    13.2 & 18:53:20.87 & $+$01:25:05.0 & 19.2 & 17.3 & 16.3 \\
  81 & $-$10.4 &    13.3 & 18:53:17.76 & $+$01:25:05.1 &      &      & 17.2 \\
  82 & $-$23.7 &    13.7 & 18:53:16.86 & $+$01:25:05.5 &      &      & 17.4 \\
  83 & $-$11.8 &    13.8 & 18:53:17.66 & $+$01:25:05.6 &      & 18.2 & 17.0 \\
  84 & $-$28.4 &    14.4 & 18:53:16.55 & $+$01:25:06.2 &      &      & 18.0 \\
  85 & $-$26.3 &    14.7 & 18:53:16.69 & $+$01:25:06.5 & 19.3 & 17.9 & 17.3 \\
  86 & $-$36.4 &    15.4 & 18:53:16.02 & $+$01:25:07.2 &      & 18.8 & 17.8 \\
  87 & $-$27.6 &    15.8 & 18:53:16.60 & $+$01:25:07.6 &      &      & 17.6 \\
  88 & $-$34.8 &    16.0 & 18:53:16.12 & $+$01:25:07.8 &      &      & 17.8 \\
  89 &  $-$5.5 &    16.7 & 18:53:18.08 & $+$01:25:08.5 &      &      & 16.6 \\
  90 &    33.9 &    17.0 & 18:53:20.71 & $+$01:25:08.7 &      &      & 17.1 \\
  91 & $-$23.5 &    18.0 & 18:53:16.88 & $+$01:25:09.8 &      &      & 15.7 \\
  92 &    30.8 &    18.1 & 18:53:20.50 & $+$01:25:09.9 & 17.0 & 16.6 & 16.2 \\
  93 &     6.0 &    18.7 & 18:53:18.84 & $+$01:25:10.5 & 17.7 & 17.1 & 16.8 \\
  94 & $-$21.7 &    18.9 & 18:53:16.99 & $+$01:25:10.7 &      & 16.9 & 15.1 \\
  95 & $-$13.2 &    19.5 & 18:53:17.56 & $+$01:25:11.3 & 19.6 & 18.0 &      \\
  96 &  $-$7.1 &    19.6 & 18:53:17.97 & $+$01:25:11.4 &      &      & 18.2 \\
  97 &    24.1 &    20.0 & 18:53:20.05 & $+$01:25:11.8 & 17.5 & 17.1 & 16.8 \\
  98 & $-$19.7 &    20.4 & 18:53:17.12 & $+$01:25:12.2 &      &      & 16.9 \\
  99 &    30.7 &    21.0 & 18:53:20.49 & $+$01:25:12.8 & 19.1 & 18.4 & 17.9 \\
      \hline
    \end{tabular}
    \end{center}
    \end{minipage}
  \end{center}
\end{table}

\addtocounter{table}{-1}
\clearpage
\begin{table}[h!t]
  \begin{center}
    \begin{minipage}{16.0cm}
    \caption{\label{T:RESULTS-NIR-4} Continued. }
    \begin{center}
%   \begin{tabular}{|r|*{2}{r|}*{5}{c|}}
    \begin{tabular}{|rrrccccc|}
      \hline
 SNB  & $\Delta$x & $\Delta$y &    RA   &   DEC   &   J   &   H   & K$^{\prime}$ \\
 $\#$ & [$\arcs$] & [$\arcs$] & (J2000) & (J2000) & [mag] & [mag] & [mag] \\
      \hline\hline
 100 &    27.3 &    22.1 & 18:53:20.26 & $+$01:25:13.9 &      &      & 18.1 \\
 101 & $-$34.6 &    22.9 & 18:53:16.14 & $+$01:25:14.6 &      &      & 17.3 \\
 102 & $-$33.1 &    23.4 & 18:53:16.24 & $+$01:25:15.1 &      &      & 17.0 \\
 103 &     3.9 &    23.7 & 18:53:18.70 & $+$01:25:15.5 & 17.3 & 16.9 & 16.6 \\
 104 & $-$22.9 &    23.9 & 18:53:16.92 & $+$01:25:15.7 &      &      & 18.5 \\
 105 & $-$28.9 &    25.5 & 18:53:16.51 & $+$01:25:17.3 &      &      & 18.0 \\
 106 & $-$11.1 &    26.3 & 18:53:17.70 & $+$01:25:18.1 &      &      & 17.7 \\
 107 &  $-$0.1 &    26.9 & 18:53:18.44 & $+$01:25:18.7 & 16.3 & 16.0 & 15.6 \\
 108 &    38.1 &    27.0 & 18:53:20.98 & $+$01:25:18.8 &      &      & 17.6 \\
 109 & $-$29.9 &    27.1 & 18:53:16.44 & $+$01:25:18.9 &      &      & 17.9 \\
 110 &  $-$8.7 &    27.4 & 18:53:17.87 & $+$01:25:19.2 &      & 18.8 & 17.3 \\
 111 & $-$30.0 &    27.8 & 18:53:16.44 & $+$01:25:19.6 &      &      & 17.8 \\
 112 & $-$17.7 &    28.0 & 18:53:17.26 & $+$01:25:19.8 & 15.1 & 14.6 & $<$~14\\
 113 &  $-$3.8 &    28.1 & 18:53:18.20 & $+$01:25:19.9 &      &      & 17.6 \\
 114 &    27.7 &    28.3 & 18:53:20.29 & $+$01:25:20.1 & 18.0 & 17.4 & 17.1 \\
 115 &    30.6 &    28.3 & 18:53:20.48 & $+$01:25:20.1 &      &      & 18.0 \\
 116 & $-$11.6 &    30.0 & 18:53:17.67 & $+$01:25:21.8 & 18.8 & 18.2 & 18.0 \\
 117 & $-$29.6 &    30.4 & 18:53:16.47 & $+$01:25:22.2 &      &      & 17.9 \\
 118 &    14.9 &    31.1 & 18:53:19.44 & $+$01:25:22.9 &      & 19.0 & 17.5 \\
 119 &    11.0 &    31.6 & 18:53:19.18 & $+$01:25:23.4 & 19.5 & 18.4 & 17.9 \\
 120 &    15.6 &    31.7 & 18:53:19.48 & $+$01:25:23.5 &      &      & 17.5 \\
 121 & $-$36.3 &    32.2 & 18:53:16.03 & $+$01:25:24.0 &      &      & 18.2 \\
 122 &  $-$2.2 &    32.8 & 18:53:18.30 & $+$01:25:24.5 & 15.4 & 15.0 & 14.7 \\
 123 &    34.6 &    33.1 & 18:53:20.75 & $+$01:25:24.9 &      & 17.9 & 16.0 \\
 124 &  $-$6.0 &    33.4 & 18:53:18.04 & $+$01:25:25.2 & 18.2 & 17.6 & 17.2 \\
 125 & $-$27.2 &    33.5 & 18:53:16.63 & $+$01:25:25.3 & 18.6 & 17.9 &      \\
 126 &    38.0 &    34.6 & 18:53:20.98 & $+$01:25:26.4 &      &      & 17.6 \\
 127 &    36.5 &    34.8 & 18:53:20.88 & $+$01:25:26.6 & 19.6 & 18.7 & 17.5 \\
 128 & $-$28.7 &    35.0 & 18:53:16.53 & $+$01:25:26.8 & 19.2 & 18.5 &      \\
 129 &     0.1 &    35.8 & 18:53:18.45 & $+$01:25:27.6 &      &      & 17.7 \\
 130 &  $-$2.1 &    36.0 & 18:53:18.31 & $+$01:25:27.8 &      &      & 17.4 \\
      \hline
    \end{tabular}
    \end{center}
    \end{minipage}
  \end{center}
\end{table}

\addtocounter{table}{-1}
\clearpage
\begin{table}[h!t]
  \begin{center}
    \begin{minipage}{16.0cm}
    \caption{\label{T:RESULTS-NIR-4} Continued. }
    \begin{center}
%   \begin{tabular}{|r|*{2}{r|}*{5}{c|}}
    \begin{tabular}{|rrrccccc|}
      \hline
 SNB  & $\Delta$x & $\Delta$y &    RA   &   DEC   &   J   &   H   & K$^{\prime}$ \\
 $\#$ & [$\arcs$] & [$\arcs$] & (J2000) & (J2000) & [mag] & [mag] & [mag] \\
      \hline\hline
 131 &  $-$9.3 &    36.1 & 18:53:17.82 & $+$01:25:27.9 &      &      & 17.6 \\
 132 &  $-$3.4 &    36.4 & 18:53:18.22 & $+$01:25:28.2 &      &      & 17.3 \\
 133 &  $-$8.0 &    36.5 & 18:53:17.91 & $+$01:25:28.3 &      &      & 17.5 \\
 134 &    32.9 &    37.0 & 18:53:20.64 & $+$01:25:28.8 & 19.5 & 18.3 & 17.6 \\
 135 & $-$25.2 &    37.5 & 18:53:16.77 & $+$01:25:29.3 &      & 18.8 & 17.6 \\
 136 &  $-$9.7 &    37.8 & 18:53:17.80 & $+$01:25:29.6 &      &      & 18.0 \\
 137 &    37.9 &    38.7 & 18:53:20.97 & $+$01:25:30.5 &      &      & 17.7 \\
 138 & $-$35.8 &    39.3 & 18:53:16.06 & $+$01:25:31.1 & 18.4 & 17.5 & 17.2 \\
 139 & $-$41.0 &    39.4 & 18:53:15.71 & $+$01:25:31.2 & 15.9 & 14.8 & $<$~14\\
 140 &    26.8 &    39.6 & 18:53:20.23 & $+$01:25:31.4 & 17.4 & 16.7 & 16.3 \\
 141 & $-$38.0 &    40.6 & 18:53:15.91 & $+$01:25:32.4 & 18.4 & 17.1 & 16.5 \\
 142 &    12.5 &    40.7 & 18:53:19.28 & $+$01:25:32.5 & 18.9 & 18.2 & 17.9 \\
 143 &    39.1 &    41.6 & 18:53:21.05 & $+$01:25:33.3 &      &      & 17.1 \\
 144 &    17.2 &    42.8 & 18:53:19.58 & $+$01:25:34.6 &      &      & 17.1 \\
 145 & $-$39.3 &    42.9 & 18:53:15.82 & $+$01:25:34.7 & 18.5 & 17.6 & 17.1 \\
 146 & $-$39.4 &    45.7 & 18:53:15.82 & $+$01:25:37.5 & 19.3 & 18.4 &      \\
      \hline
    \end{tabular}
    \end{center}
    \begin{center}
        Notes: Typical photometric errors are
               $\sigma_{\rm J}$\,$\sim$\,1$\fmag$2\,$\pm$\,0$\fmag$5,
               $\sigma_{\rm H}$\,$\sim$\,1$\fmag$4\,$\pm$\,0$\fmag$6 and
               $\sigma_{\rm K^{\prime}}$\,$\sim$\,1$\fmag$6\,$\pm$\,0$\fmag$6.
    \end{center}
    \end{minipage}
  \end{center}
\end{table}

\normalsize
\clearpage
%table 2:
\small
\begin{table}[h]
%% \caption[]{Derived Parameters for ionized gas detected at 6\,cm
%% wavelength}  
\caption[]{\label{T:RESULTS-CM} Derived Parameters for ionized gas 
detected at 6\,cm wavelength}

\smallskip
\begin{tabular}{|lccccccccc|}
\hline
%%           & $S\nu^\dagger$ & $\Delta s$ & T$_b$ & $\tau_\nu$ & EM  
          & $S_{\nu}^\dagger$ & $\Delta s$ & T$_b$ & $\tau_\nu$ & EM  
            & $n_e$    & U             & Log$N_L$  & Spectral \\ 
Source    & (mJy)  & (pc)       & (K)   &            & (pc cm$^{-6}$)
          & (cm$^{-3}$)& (pc cm$^{-2}$)& ($s^{-1}$)& Type \\ 

\hline
\hline
% src        S_nu   s        Tb   tau                  EM 
G34.4\,MM   &0.7   &$<0.21$ &2.4 &$2.4 \times 10^{-4}$ &$2.1 \times 10^{5}$ 
%            ne        U    logN_L  spty
            &$> 10^3$ &2.3 &44.76  &B2 \\
G34.4$+$0.23&9.0   &0.12$^{\dagger\dagger}$ &31.0&$3.1 \times 10^{-3}$ 
	    &$2.7 \times 10^{5}$  
            &$> 10^3$ &5.3 &45.87  &B0.5 \\

\hline
\end{tabular}

\vspace{.1in}
~{\small $^\dagger$~~Flux densities measured in the primary beam
        corrected image.} \\
~~{\small $^{\dagger\dagger}$~~Deconvolved size from Molinari et
al. (1998).}  
\end{table}
\normalsize
\clearpage
\vspace{2cm}
% numbers verified on 12sep02
% Table 3: 
\begin{table}[h]
%% \caption{{\htco}(J=1--0) Mass and Column Density Estimates}
\caption{\label{T:RESULTS-MM}{\htco}(J=1--0) Mass and Column Density Estimates}

\smallskip
\begin{tabular}{|lcccc|}
\hline 
&                           &              &Ave {\htco}    &Inferred  \\
& Peak Position             &{\htco} mass  &column density &{\h} Mass \\
& (h~m~s)~~~(\dec{~}{~}{~}) &(M\sun)       &(cm$^{-2}$)    &(M\sun)   \\
\hline 
\hline 
G34.4~MM core~~~~~~& 18 53 18.00~~+01 25 24.9
              &$9.3 \times 10^{-8}$ &$1.7 \times 10^{14}$ &$4 \times 10^{3}$\\
Southern core      & 18 53 18.58~~+01 24 50.5
              &$1.3 \times 10^{-7}$ &$2.4 \times 10^{14}$ &$5 \times 10^{3}$\\
&             &                     &                     &          \\
Total$^\dagger$    & \nodata
              &$1.1 \times 10^{-6}$ &$1.3 \times 10^{14}$ &$5 \times 10^{4}$\\
\hline 
\end{tabular}

\vspace{.1in}
~~{\small $^\dagger$~~Total includes core emission and diffuse
component. } 
\end{table}

\clearpage
\begin{center}  {\bf Figure Captions}  \end{center}

\noindent
{\bf Figure~1.~~} 
Millimeter (thin lines) \& centimeter (thick
lines) continuum emission toward IRAS\,18507.
The 3~mm image has an RMS of 3.3\mjyb\ and 
a peak flux density of 32.0\mjyb\ at the position of G34.4 MM.
Contours (thin lines) are plotted at $\pm$ 3, 5, 7, \& 9 $\sigma$.
The 6~cm image has an RMS of 0.19\mjyb\ and a peak flux density of
8.72\mjyb.  Contours (thick lines) are plotted at $\pm$ 3, 10, 20, 30,
40 $\sigma$.  Synthesized beams for both observations are plotted in
the lower right. A scale size of 0.4 pc is represented by a bar in 
the upper left corner.  

\noindent
{\bf Figure~2.~~}  
H$^{13}$CO$^{+}$ emission. 
The top left panel shows integrated H$^{13}$CO$^{+}$ emission 
(moment 0) between 55.25\kms\ and 62.15\kms\ ($v_{LSR} = 57$\kms). 
The RMS in the image is 230\mjybkms; contours are plotted at $\pm 3,
5, 7, 9, \& 11 \sigma$.  The remaining panels show the H$^{13}$CO$^{+}$ 
channel maps at 1.7\kms\ spectral resolution.  The central velocity is
indicated in the upper left of each panel.  The channel RMS is
50\mjyb\ and contours are plotted at $\pm 3, 5, 7, 9, \& 11
\sigma$. The lower right panel shows the synthesized beam in the
bottom right corner ($5.34'' \times 4.51''$ at P.A. $-43.8^\circ$) and
a scale size of 0.55 pc.  Plus symbols represent the locations of the
6~cm continuum peak and G34.4~MM.

\noindent
{\bf Figure~3.~~}  
Near-infrared images of the IRAS\,18507 star forming region in the broad
bands J (top panel; 1.25\,$\mu$m), H (center panel; 1.65\,$\mu$m), and
K$'$ (bottom panel; 2.10\,$\mu$m).  Plus symbols represent the
locations of the 6~cm continuum peak and G34.4~MM.  The location of
the near-infrared source $\#$54 is shown in the center panel.  

\noindent
{\bf Figure~4.~~} The K$'$ image of Fig 3. shown in grey-scale
compared with a map of the integrated {\htco} emission between
55.25\kms\ and 62.15\kms\ (from top left panel in Fig. 2).  Plus
symbols represent the locations of the 6~cm continuum peak and
G34.4~MM; source $\#$54 also identified.  The location of IRAS
18507+0121 is illustrated by the cross centered at $\alpha =$
\ra{18}{53}{17.42} ~ $\delta =$ \dec{+01}{24}{54.5}~(J2000).  The
length and orientation of the symbol represents the positional
accuracy.

\noindent
{\bf Figure~5.~~} (H$-$K$^{\prime}$)--(J$-$H) diagram of the detected
near-infrared sources without applying any correction for possible
foreground extinction.  The location of source $\#$54 is emphasized by
an asterisk.  Typical loci of dwarfs and giants are marked by the
thick and thin lines.  Further explanations are given in the text.

\noindent
{\bf Figure~6.~~} M$_{\rm J}$--(J$-$K$^{\prime}$) diagram of the
detected near-infrared sources, again without applying any correction
for possible foreground extinction.  The location of source $\#$54 is
highlighted by an asterisk.  Loci of stars on the main sequence (thick
line) and on the pre-main sequence at ages of 0.3\,Myr and 3\,Myr (two
dashed lines) are outlined.

%%%%%%%%%%%%%%%%%%%%%%%%%%%%%%%%%%%%%%%%%%%%%%%%%%%%%%%%%%%%%%%%%%%%%%%%%%%%
\end{document}